\documentclass[preprint,prd,aps,nofootinbib]{revtex4}
\usepackage{graphicx}
\usepackage{bm}
\usepackage{amssymb}
\usepackage{amsmath}
\usepackage{euscript}

\begin{document}

\title{Large Scale Evolution of Premixed Flames}

\author{Kirill~A.~Kazakov\thanks{E-mail: $kirill@theor.phys.msu.ru$}}

\affiliation{Department of Theoretical Physics, Physics Faculty,
Moscow State University, $119899$, Moscow, Russian Federation}

\begin{abstract}
The influence of the small scale ``cellular'' structure of
premixed flames on their evolution at larger scales is
investigated. A procedure of the space-time averaging of the flow
variables over flame cells is introduced. It is proved that to the
leading order in the flame front thickness, the form of dynamical
equations for the averaged gas velocity and pressure, as well as
of jump conditions for these quantities at the flame front, is the
same as in the case of a zero-thickness flame propagating in an
ideal fluid at constant velocity with respect to the fuel, equal
to the adiabatic velocity of a plane flame times a factor
describing increase of the flame front length due to the local
front wrinkling. As an application, the large scale evolution of a
flame in the gravitational field is investigated. A weakly
nonlinear non-stationary equation for the averaged flame front
position is derived. It is found that the leading nonlinear
gravitational effects stabilize the flame propagating in the
direction of the field. The resulting stationary flame
configurations are determined analytically.
\end{abstract}

\maketitle

\section{introduction}

Propagation of plane flames in gaseous mixtures is well known to
be unstable. An efficient way of investigating this instability is
to consider the flame front as a surface of discontinuity,
expanding all quantities of interest in powers of $L_{\rm
f}/\lambda,$ where $L_{\rm f}$ is the flame front thickness, and
$\lambda$ characteristic length scale of a flame perturbation. The
leading term of the perturbation growth rate expansion has the
form
\begin{eqnarray}\label{sigma0}
\sigma = c_0\frac{U_{\rm f}}{\lambda}
\end{eqnarray}
\noindent where $U_{\rm f}$ denotes an adiabatic velocity of a
plane flame front with respect to the fuel, and $c_0=c_0(\theta)$
a function of the gas expansion coefficient $\theta$ defined as
the ratio of the fuel density ($\rho_{\rm u}$) and the density of
burnt matter ($\rho_{\rm b}$), $\theta = \rho_{\rm u}/\rho_{\rm b
}>1.$ According to Refs.~\cite{landau,darrieus}, $c_0(\theta)$ has
positive values for all $\theta,$ implying an unconditional
instability of zero-thickness flames, the Landau-Darrieus (LD)
instability. In the next order in $L_{\rm f}/\lambda,$ account of
the transport processes inside the flame front modifies
Eq.~(\ref{sigma0}) to
\begin{eqnarray}\label{sigma1}
\sigma = c_0\frac{U_{\rm f}}{\lambda}\left(1 - c_1\frac{L_{\rm
f}}{\lambda}\right)\,,
\end{eqnarray}
\noindent where $c_1$ depends on $\theta$ as well as on the ratio
of the heat and mass diffusivities (the Lewis number)
\cite{markstein,pelce,matalon}. The product $c_1L_{\rm f}\equiv
\lambda_{\rm c},$ the so-called cut-off wavelength, is the short
wavelength limit of unstable perturbations. By the order of
magnitude, $\lambda_{\rm c}$ represents also the characteristic
length of the so-called cellular structure of the flame front,
which is formed eventually as a result of the nonlinear flame
stabilization. For many flames of practical interest, $c_1 = 15 -
20.$ It is the fact that $L_{\rm f}$ is relatively small in
comparison with $\lambda_{\rm c}$ which underlies the above point
of view on flame dynamics.

Consider an arbitrary initially smooth front configuration. As a
result of the rapid growth of the unstable flame perturbations
with wavelengths $\sim\lambda_{\rm c},$ the flame front becomes
corrugated within the time interval
\begin{eqnarray}\label{ctime}
\Delta t \sim \frac{\lambda_{\rm c}}{U_{\rm f}}\,.
\end{eqnarray}
\noindent Dynamics of the short wavelength modes are mainly
determined by the transport processes inside the flame front, and
are affected only slightly by the large scale flow. One can say
that the small scale cellular structure of the flame front
develops on the ``background'' of its smooth large scale
configuration (see Fig.~1). It follows all developments of the
background, Eq.~(\ref{ctime}) playing the role of the
characteristic time of cell adaptation to the large scale front
evolution.

\begin{figure}
\scalebox{0.9}{\includegraphics{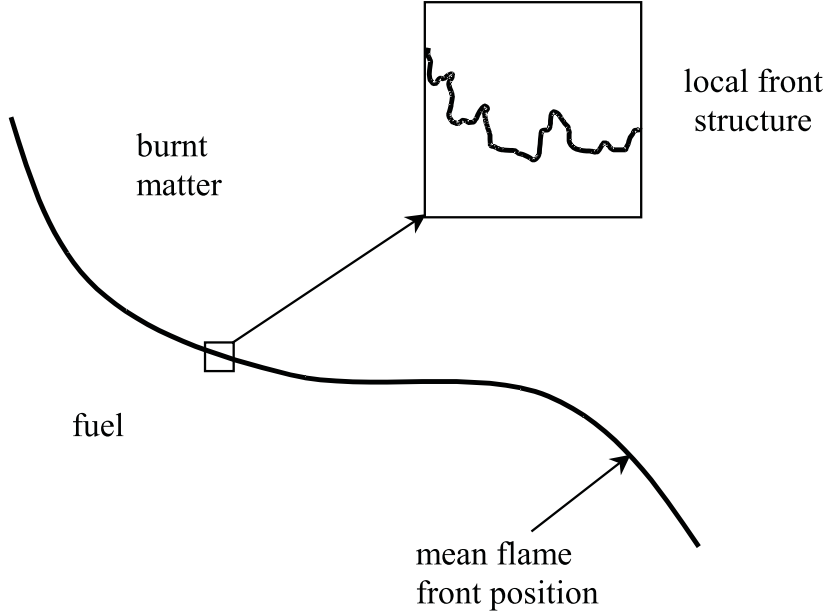}} \label{fig1}
\caption{Schematic representation of the scale separation.}
\end{figure}

In practice, it is the large scale evolution of the flame front,
rather than its exact local structure, which is often of the main
concern. In this respect, an important question arises about the
reverse influence of the flame cellular structure on the front
evolution at scales much larger than $\lambda_{\rm c}.$ More
precisely, one can state the problem as follows. Imagine that we
have smeared the small scale rapid variations of all relevant
quantities by averaging them over many flame cells. Then the
question is what equations governing dynamics of the averages are.

In connection with the above statement of the problem it should be
noted that the exact cellular structure of flames is actually
unknown. This is because the process of cell formation is
essentially nonlinear, in the sense that it cannot be treated
perturbatively in principle, which is the main reason of lack of
its theoretical description. Only in the case $\theta \to 1,$
which is practically irrelevant, can this structure be determined
analytically \cite{henon,siv1,sivclav,kazakov1,kazakov2}. The
question of principle, therefore, is to what extent the large
scale dynamics of averages depend on the exact local flame
structure in the regime of fully developed LD-instability.

The main purpose of the present paper is to show that to the
leading order in the ratio of $\lambda_{\rm c}$ to the
characteristic length of the problem ($L_0$), dynamics of the
averaged quantities are actually independent of particularities of
the local flame structure. The latter determines essentially only
one parameter characterizing the large scale evolution -- the
effective normal velocity of the flame front.

Perhaps, it is worth to explain the essence of the problem in a
little bit more detail. The above point of view on the flame
propagation is based on the possibility to separate the local
cellular dynamics from the large scale evolution of the
background. This possibility is underlined by the following common
property of the transport processes. From the mathematical point
of view, all these processes are of higher differential order than
those governing dynamics of an ideal fluid. Therefore, their
relative role increases at smaller scales. In particular, in the
limit $L_{\rm f}\to 0,$ cell formation is completely determined by
the transport processes inside the flame front. On the contrary,
the role of these processes at scales $L\gg L_{\rm f}$ is
relatively small. It should be fully realized, however, that this
reasoning is inherently linear. It tacitly assumes that if every
quantity of interest, say $A,$ is represented as a sum of its
averaged value $\langle A\rangle \equiv A_0$ and the small scale
fluctuation $A_1,$ then the dynamics of $A_0$'s can be determined
solely in terms of $A_0$'s themselves. Because of the high
nonlinearity of basic equations governing the flame propagation,
this assumption is far from being self-evident.  For instance,
averaging of a cubic combination of $A$'s gives rise to a term
$\langle A_1^2\rangle A_0$ comparable with $A_0^3,$ since the
small and large scale parts of the flow variables are generally of
the same order of magnitude. Clearly, equations for $A_0$'s
involving such terms would not be of great value, since the local
flame dynamics, and therefore, the coefficients $\langle
A_1^2\rangle,$ are unknown. The main result of the present work is
the proof that such terms actually do not arise in the leading
order with respect to $\lambda_{\rm c}/L_0.$ One can say that the
governing equations for the quantities $A_0$ and $A_1$ decouple
from each other. The proof consists of two parts corresponding to
decoupling of the flow equations in the bulk, and decoupling of
the jump conditions at the flame front, given in
Secs.~\ref{eqdecoupling} and \ref{jdecoupling}, respectively. As
an application of this result, the problem of nonlinear front
stabilization in a gravitational field will be considered in
Sec.~\ref{example}. The results obtained are discussed in
Sec.~\ref{conclude}.

\section{The decoupling theorem}\label{theorem}

\subsection{The averaging procedure}\label{aprocedure}

Let us begin with the precise formulation of the averaging
procedure. Denote $L_0$ the characteristic length of the problem
in question. For instance, $L_0$ can be the tube width, in the
case of a flame propagating in a tube, or be related to an
external field acting on the system. In practice, this length
largely exceeds the flame cell size, $$L_0\ggg \lambda_{\rm
c}\,.$$ Assuming this, let us choose a length $L$ satisfying
\begin{eqnarray}
\lambda_{\rm c} \ll L \ll L_0\,. \nonumber
\end{eqnarray}
\noindent Analogously, denoting the characteristic time interval
by $T_0,$ and noting that $$T_0 \sim \frac{L_0}{U_{\rm f}}\,,$$ we
can choose $T\sim L/U_{\rm f}$ such that
$$\frac{\lambda_{\rm c}}{U_{\rm f}}\ll T\ll T_0\,.$$
Given a function $A(\bm{x},t),$ we define its space-time average
over $\{\bm{x},t: \bm{x}\in (\bm{x}_0,\bm{x}_0 + \Delta\bm{x}),$
$\Delta x_i = L,$ $~t\in (t_0,t_0+T)\}$
\begin{eqnarray}\label{average}
\langle A \rangle = \frac{1}{L^3 T} \int\limits_T dt
\int\limits_{V}d^3\bm{x}~A(\bm{x},t) \equiv A_0(\bm{x}_0,t_0)\,.
\end{eqnarray}
\noindent By the definition, $\langle A\rangle$ varies noticeably
over space distances $\|\Delta \bm{x}_0\| \sim L_0,$ and time
intervals $\Delta t_0 \sim T_0.$ The function $A$ thus turns out
to be decomposed into two parts corresponding to the two scales,
$L_0$ and $\lambda_{\rm c}:$
$$A = A_0 + A_1, \qquad \langle A_1 \rangle = 0.$$
As was mentioned in the Introduction, flame dynamics can be
analyzed in the framework of the power expansion with respect to
the small ratio $L_{\rm f}/L_0\equiv\varepsilon$ (or equivalently,
with respect to $\lambda_{\rm c}/L_0,$ since $\lambda_{\rm c} =
O(L_{\rm f})$). Thus, we write $A_0$ and $A_1$ as follows
$$A_0 = A_0^{(0)} + \varepsilon A_0^{(1)} + \cdot\cdot\cdot\,,
\quad A_1 = A_1^{(0)} + \varepsilon A_1^{(1)} +
\cdot\cdot\cdot\,,$$ dots denoting terms of higher order in
$\varepsilon.$ In this notation, the large scale flame dynamics in
zero order approximation with respect to $\varepsilon$ are
described by the quantities $A_0^{(0)}.$ Our main purpose below
will be to investigate coupling between $A_0^{(0)}$ and
$A_1^{(0)},$ $A_1^{(1)},$ etc., and to obtain effective equations
governing dynamics of $A_0^{(0)}.$ Accordingly, all quantities
will be measured in units relevant to the large scale dynamics.
Namely, space coordinates $\bm{x}$ and time $t$ are assumed to be
normalized on $L_0$ and $L_0/U_{\rm f},$ respectively.
Furthermore, $U_{\rm f}$ will be taken as the unit of gas velocity
$\bm{v},$ while $\rho_{\rm u} U_{\rm f }^2$ as the unit of gas
pressure $p\,.$ For future reference, let us write down their
expansions explicitly
\begin{eqnarray}\label{01expansion}&&
\bm{v} = \bm{v}_0 + \bm{v}_1\,, \quad p = p_0 + p_1\,, \\&&
\bm{v}_0 = \bm{v}_0^{(0)} + \varepsilon\bm{v}_0^{(1)} +
\cdot\cdot\cdot\,, \quad \bm{v}_1 = \bm{v}_1^{(0)} +
\varepsilon\bm{v}_1^{(1)} + \cdot\cdot\cdot\,,
\label{vexpansion}\\&& p_0 = p_0^{(0)} + \varepsilon p_{0}^{(1)} +
\cdot\cdot\cdot\,, \quad p_1 = p_1^{(0)} + \varepsilon p_1^{(1)} +
\cdot\cdot\cdot\,. \label{pexpansion}
\end{eqnarray}
\noindent Within our choice of units, we have the following
estimates
\begin{eqnarray}\label{orders0}
\bm{v}_0 &=& O(1), \quad \bm{v}_1 = O(1), \quad p_0 = O(1),
\quad p_1 = O(1),\\
\frac{\partial v_{0i}}{\partial x_k} &=& O(1), \quad
\frac{\partial v_{0i}}{\partial t} = O(1), \quad \frac{\partial
p_0}{\partial x_k} = O(1), \\
\frac{\partial v_{1i}}{\partial x_k} &=&
O\left(\frac{1}{\varepsilon}\right), \quad\frac{\partial
v_{1i}}{\partial t} = O\left(\frac{1}{\varepsilon}\right),
\quad\frac{\partial p_1}{\partial x_k} =
O\left(\frac{1}{\varepsilon}\right),\\
\quad\frac{\partial^2 v_{0i}}{\partial x_k\partial x_l} &=&
O(1)\,, \quad\frac{\partial^2 v_{1i}}{\partial x_k\partial x_l} =
O\left(\frac{1}{\varepsilon^2}\right)\,, \quad i,k,l =
1,2,3.\label{orders2}
\end{eqnarray}
\noindent Let us now proceed to the examination of flame dynamics
in terms of $\bm{v}_{0,1}^{(0,1)}, p_{0,1}^{(0,1)}.$

The main result concerning the large scale flame dynamics which
will be proved below can be expressed in the form of the following

\noindent {\it Decoupling theorem:} The large scale dynamics of a
flame are unaffected by its local cellular structure up to a
rescaling. More precisely, the form of dynamical equations for
$\bm{v}_0^{(0)},$ $p_0^{(0)},$ as well as of jump conditions for
these quantities at the flame front, is the same as in the case of
zero-thickness flame propagating in an ideal fluid at constant
speed $\mathfrak{U}~U_{\rm f}$ with respect to the fuel, the
number $\mathfrak{U}>1$ describing the flame front length increase
due to the local front wrinkling.

The {\it proof} consists of two parts corresponding to decoupling
of the flow equations in the bulk, and decoupling of the jump
conditions at the flame front, presented in
Secs.~\ref{eqdecoupling} and \ref{jdecoupling}, respectively.

\subsection{Decoupling of dynamical equations}\label{eqdecoupling}

For definiteness, we will assume in what follows that external
field acting on the system is the gravitational field, denoting
its strength by $\bm{g}.$ Accordingly, $L_0$ will be identified
with the characteristic length associated with this
field:\footnotemark \footnotetext{If the gravitational field is
not homogeneous, it is assumed to vary noticeably over distances
larger than~$L_0.$}
\begin{eqnarray}\label{l0}
L_0 = \frac{U^2_{\rm f}}{\|\bm{g}\|}\,.
\end{eqnarray}
\noindent Then the dimensionless velocity and pressure fields obey
the following equations in the bulk
\begin{eqnarray}\label{cont}
{\rm div}~\bm{v} &=& 0\,, \\
\frac{\partial\bm{v}}{\partial t} + (\bm{v}\bm\nabla)\bm{v} &=&
-\frac{1}{\varrho}\bm\nabla p + \bm{G} + \varepsilon
Pr\triangle\bm{v}\,, \label{euler}
\end{eqnarray}
\noindent where $$\bm{G} = \frac{\bm{g}L_0}{U_{\rm f}^2}\,, \quad
\left\|\bm{G}\right\| = 1,$$ $\varrho$ is the fluid density
normalized on the fuel density $\rho_{\rm u},$ and $Pr$ the
Prandtl number representing the ratio of viscous and thermal
diffusivities, $Pr = \nu/\chi.$

Substituting expansions (\ref{01expansion})--(\ref{pexpansion})
into Eqs.~(\ref{cont}), (\ref{euler}), taking into account the
estimates (\ref{orders0})--(\ref{orders2}), and extracting
$O(1/\varepsilon)$ terms yields
\begin{eqnarray}\label{flow0}
{\rm div}~\bm{v}_1^{(0)} &=& 0\,,\\
\frac{\partial\bm{v}_1^{(0)}}{\partial t} +
\left(\left[\bm{v}_0^{(0)} +
\bm{v}_1^{(0)}\right]\bm\nabla\right)\bm{v}_1^{(0)} &=&
-\frac{1}{\varrho}\bm\nabla p_1^{(0)} + \varepsilon
Pr\triangle\bm{v}_1^{(0)}\,. \label{flow0euler}
\end{eqnarray}
\noindent Next, collecting $O(1)$ terms gives
\begin{eqnarray}\label{flow1cont}
{\rm div}~\bm{v}_0^{(0)} &=& 0\,, \\
\frac{\partial\bm{v}_0^{(0)}}{\partial t} +
\varepsilon\frac{\partial\bm{v}_1^{(1)}}{\partial t} +
\left(\left[\bm{v}_0^{(0)} +
\bm{v}_1^{(0)}\right]\bm\nabla\right)\bm{v}_0^{(0)} &+&
\varepsilon \left(\left[\bm{v}_0^{(0)} +
\bm{v}_1^{(0)}\right]\bm\nabla\right)\bm{v}_1^{(1)} \nonumber\\
= - \frac{1}{\varrho}\bm\nabla p_0^{(0)} &-&
\frac{\varepsilon}{\varrho}\bm\nabla p_1^{(1)} + \bm{G} +
\varepsilon^2 Pr\triangle\bm{v}_1^{(1)}\,. \label{flow1euler}
\end{eqnarray}
\noindent Equation (\ref{flow1euler}) involves both slowly and
rapidly varying terms. The slowly varying part of this equation,
determining dynamics of the fields $\bm{v}_0^{(0)},$ $p_0^{(0)},$
can be separated out by averaging it according to
Eq.~(\ref{average}). Under this operation, all terms linear in
$\bm{v}_1, p_1$ give rise to $o(1)$ contribution. For instance,
\begin{eqnarray}\label{aveuler}
\varepsilon\left\langle\frac{\partial\bm{v}_1^{(1)}}{\partial
t}\right\rangle = \frac{\varepsilon}{L^3
T}\int\limits_{V}d^3\bm{x}\left.\bm{v}_1^{(1)}\right|_{t_0}^{t_0 +
T} = \frac{L_{\rm f}}{L_0} O\left(\frac{T_0}{T}\right) =
O\left(\frac{L_{\rm f }}{L}\right) = o(1)\,,
\end{eqnarray}\noindent in view of the estimates (\ref{orders0}),
and the choice of $L,T.$ The same argument applies to
$(\bm{v}_0^{(0)}\bm\nabla)\bm{v}_1^{(1)},$ as well as to the
second and fourth terms in the right hand side of
Eq.~(\ref{flow1euler}). Furthermore,
$$\left\langle\left(\bm{v}_1^{(0)}\bm\nabla \right)\bm{v}_0^{(0)}
\right\rangle \equiv 0$$ according to the definition of
$\bm{v}_1.$ Finally, contribution of the last term in the left
hand side of (\ref{flow1euler}) also is $o(1).$ Indeed,
integrating by parts and taking into account Eq.~(\ref{flow0}), we
have
\begin{eqnarray}\label{surface}
\left\langle\left(\bm{v}_1^{(0)}\bm\nabla\right)
\bm{v}_1^{(1)}\right\rangle = \frac{1}{L^3 T}\int\limits_T
dt\int\limits_{V}d^3\bm{x}~\left(\bm{v}_1^{(0)}
\bm\nabla\right)\bm{v}_1^{(1)} = \frac{1}{L^3 T}\int\limits_T
dt\int\limits_{S}\left(d\bm{s}~\bm{v}_1^{(0)}\right)\bm{v}_1^{(1)}\,,
\end{eqnarray} \noindent where $S$ is the surface of the cube
$V = \{\bm{x}: \bm{x}\in (\bm{x}_0,\bm{x}_0 + \Delta\bm{x}),$
$\Delta x_i = L\},$ $d\bm{s}$ being its element. Using
Eqs.~(\ref{orders0}), the right hand side of Eq.~(\ref{surface})
is estimated as $O(L_0/L).$ Hence,
$$\varepsilon\left\langle\left(\bm{v}_1^{(0)}\bm\nabla\right)
\bm{v}_1^{(1)}\right\rangle = \frac{L_{\rm
f}}{L_0}O\left(\frac{L_0}{L}\right) = o(1)\,.$$

Thus, Eq.~(\ref{flow1euler}) reduces upon averaging to the
ordinary Euler equation for the functions $\bm{v}_0^{(0)},$
$p_0^{(0)}$
\begin{eqnarray}\label{aveuler1}
\frac{\partial\bm{v}_0^{(0)}}{\partial t} + \left(\bm{v}_0^{(0)}
\bm\nabla\right)\bm{v}_0^{(0)} = - \frac{1}{\varrho}\bm\nabla
p_0^{(0)} + \bm{G} + o(1)\,,
\end{eqnarray} \noindent
which proves the first part of the decoupling theorem. It is worth
of mentioning that the large scale flow dynamics in the bulk turn
out to be ideal at zeroth order in $\varepsilon.$

In connection with Eqs.~(\ref{flow0}), (\ref{flow0euler}) the
following circumstance should be emphasized. These equations
describe bulk dynamics of the small scale parts of the flow
variables at zeroth order in $L_{\rm f}/L_0,$ {\it i.e.,} when the
influence of the large scale flows on the flame cellular structure
is completely neglected. This might seem to be in contradiction
with the structure of Eq.~(\ref{flow0euler}), because it involves
$\bm{v}_0^{(0)}$ explicitly. However, the functions
$\bm{v}_1^{(0)}(\bm{x},t),$ $p_1^{(0)}(\bm{x},t),$ satisfying
Eq.~(\ref{flow0euler}) in a given space-time region $\{\bm{x},t:
\bm{x}\in (\bm{x}_0,\bm{x}_0 + \Delta\bm{x}),$ $\Delta x_i = L,$
$~t\in (t_0,t_0+T)\},$ can be written as
$$\bm{v}_1^{(0)}(\bm{x},t) = \tilde{\bm{v}}_1^{(0)}
\left(\bm{x} - \bm{v}^{(0)}_0(\bm{x}_0,t_0)t, t\right)\,, \quad
p_1^{(0)}(\bm{x},t) = \tilde{p}_1^{(0)}\left(\bm{x} -
\bm{v}^{(0)}_0(\bm{x}_0,t_0)t, t\right)\,,$$ where
$\tilde{\bm{v}}_1^{(0)}\,,$ $\tilde{p}_1^{(0)}$ satisfy
\begin{eqnarray}
\frac{\partial\tilde{\bm{v}}_1^{(0)}}{\partial t} +
\left(\tilde{\bm{v}}_1^{(0)}\bm\nabla\right)\tilde{\bm{v}}_1^{(0)}
&=& -\frac{1}{\varrho}\bm\nabla \tilde{p}_1^{(0)} + \varepsilon
Pr\triangle\tilde{\bm{v}}_1^{(0)}\,. \nonumber
\end{eqnarray}
\noindent In other words, the role of $\bm{v}_0^{(0)}$ in
Eq.~(\ref{flow0euler}) is purely kinematical: it describes the
large scale ``drift'' of the flame cellular structure.

\subsection{Decoupling of jump conditions}\label{jdecoupling}

The proof of decoupling of the jump conditions is more
complicated, since this is the place where the transport processes
inside the flame front come into play. These conditions express
the conservation of energy and momentum across the flame front.
For freely propagating flames, and within the accuracy of the
first order in the small front thickness, they were derived in the
most general form in Ref.~\cite{matalon}. To take into account the
influence of gravity, it is sufficient to note that the bulk
equations (\ref{euler}) can be rendered formally free by
substituting $p = \tilde{p} + \varrho(\bm{G}\bm{x}).$ However,
gravity reappears through the jump conditions at the flame front.
On the other hand, the influence of gravity on gas dynamics inside
the flame front is small in comparison with the transport effects;
their relative value is known to be given by the inverse Froude
number $Fr^{-1} = L_{\rm f}\|\bm{g}\|/U^2_{\rm f} = \varepsilon.$
To the leading order in $\varepsilon,$ therefore, contribution of
the gravitational field to the jump conditions is the same as in
the case of a zero-thickness flame.

For simplicity, we will consider two-dimensional (2D) case,
assuming also that the Lewis number is equal to unity. No
assumption is made concerning the incoming flow, except that its
characteristic length $\tilde{L}\ge L_0.$ Let the flame front
position be described by an equation $z=f(x,t),$ where the
Cartesian coordinates $(x,z)$ are scaled on $L_0,$ and chosen so
that $z$-axis is parallel to $\bm{G}.$ The $x$- and $z$-components
of the flow velocity will be denoted by $w$ and $u,$ $\bm{v} =
(w,u).$ We also introduce the unit vector $\bm{\tau}$ tangential
to the flame front, and $\bm{n}$ orthogonal to it (pointing to the
burnt matter). In components,
\begin{eqnarray}\label{taun}
\bm{\tau} = \left(\frac{1}{N}\,,\frac{\partial f/\partial
x}{N}\right)\,, \quad \bm{n} =  \left(- \frac{\partial f/\partial
x}{N}\,,\frac{1}{N}\right)\,, \quad N \equiv \sqrt{1 +
\left(\frac{\partial f}{\partial x }\right)^2}\,.
\end{eqnarray}\noindent
Rewriting Eqs.~(5.32)--(5.43) of \cite{matalon} in this notation
for the 2D case, and taking into account the contribution of the
archimedean force to the pressure jump yields
\begin{eqnarray}
(\bm{v}_+ \bm{n}) - (\bm{v}_- \bm{n}) &=&
(\theta - 1)\,,\label{conserv1}\\
(\bm{v}_+ \bm{\tau}) - (\bm{v}_- \bm{\tau}) &=&
\varepsilon\left(\ln\theta + (\theta -
1)Pr\right)\frac{1}{N}\left(\hat{D}w_{-} + \frac{\partial f
}{\partial x}\hat{D}u_{-} + \frac{1}{N}\hat{D}\frac{\partial
f}{\partial x} \right)\,,\label{conserv2}\\
p_+ - p_- &=& \tilde{p}_+  - \tilde{p}_- + \left(- \varrho G_z z
\right)_+ - \left(- \varrho G_z z\right)_- \nonumber\\
&=& - (\theta - 1) - \frac{\theta - 1 }{\theta} G f + \varepsilon
(\theta - 1)\frac{\partial}{\partial
x}\left(\frac{1}{N}\frac{\partial f}{\partial x}\right)
\nonumber\\ &+&
\frac{\varepsilon\ln\theta}{N}\left(\frac{\partial^2 f}{\partial
 t^2} + 2w_{-}\frac{\partial^2 f}{\partial t
\partial x} + w_{-}^2 \frac{\partial^2 f}{\partial x^2}
+ 2 \hat{D}N - \frac{1}{N}\frac{\partial f}{\partial
x}\frac{\partial N }{\partial x}\right)\,,\label{conserv3}
\end{eqnarray}
\noindent where
$$\hat{D}\equiv \frac{\partial}{\partial t} +
\left(w_{-} + \frac{1}{N}\frac{\partial f}{\partial
x}\right)\frac{\partial}{\partial x}\ , \quad G \equiv - G_z\,,$$
and the subscripts ``$-$'' and ``$+$'' mean that the corresponding
quantity is calculated for $z = f(x,t)-0$ and $z = f(x,t) + 0,$
respectively.

Finally, to complete the system of hydrodynamic equations and jump
conditions, one needs an expression for the local burning rate.
This expression, the so-called evolution equation, has the
following form (Cf.~Eq.~(6.1) in Ref.~\cite{matalon})
\begin{eqnarray}\label{evolution}&& (\bm{v}_- \bm{n})
- \frac{1}{N}\frac{\partial f}{\partial t} = 1 -
\frac{\varepsilon\theta\ln\theta}{N(\theta -
1)}\left(\frac{\partial N}{\partial t} + \frac{\partial}{\partial
x}(N w_{-}) + \frac{\partial^2 f }{\partial x^2}\right)\,.
\end{eqnarray}
\noindent At this point, it is worth to make the following comment
on the meaning of the above asymptotic relations. Equations
(\ref{conserv1})--(\ref{evolution}) were derived in
Ref.~\cite{matalon} under assumption that the terms in the right
hand sides of these equations, proportional to $\varepsilon,$ are
small, which is only true if the gas flow is characterized by a
length scale much larger than the flame front thickness. However,
the rapidly developing LD-instability makes any smooth flame
configuration highly corrugated within the time interval of the
order (\ref{ctime}). As a result, the $\varepsilon$-terms turn out
to be of the order $L_{\rm f}/\lambda_{\rm c} = O(1).$ Similarly,
account of the $\varepsilon^2$-corrections in the above equations
would give rise to terms of the order $L^2_{\rm f}/\lambda^2_{\rm
c} = O(1)$ {\it etc.}, questioning thereby validity of the small
$\varepsilon$-expansion. However, it was mentioned in the
Introduction that in practice, the cut-off wavelength
$\lambda_{\rm c}$ is noticeably larger than the flame front
thickness $L_{\rm f}.$ Thus, in the regime of fully developed
LD-instability, the right hand sides of
Eqs.~(\ref{conserv1})--(\ref{evolution}) are to be considered the
leading order terms of the asymptotic expansion in powers of
$L_{\rm f}/\lambda_{\rm c} = 1/c_1\,,$ rather than $L_{\rm f}/L_0
= \varepsilon.$ On the contrary, $\varepsilon$ is the true
parameter of the power expansions (\ref{vexpansion}),
(\ref{pexpansion}), which determines the relative order of
successive terms in these expansions.

In order to extract from Eqs.~(\ref{conserv1})--(\ref{conserv3})
jump conditions for the quantities $\bm{v}_0^{(0)},$ $p_0^{(0)},$
we need to introduce an auxiliary operation of averaging along the
front. Let a quantity $A$ be defined on the flame front, i.e., for
$\{x,z,t: z = f(x,t)\}.$ Given a point $x_0,$ choose $\Delta x =
\Delta x(x_0)$ such that the front length $\mathfrak{L}(t)$
between the points $(x_0,f(x_0,t))$ and $(x_0,f(x_0 + \Delta
x,t))$ satisfies
$$\lambda_{\rm c}\ll \mathfrak{L}(t) \ll L_0, \quad \mathfrak{L}(t) = O(L)$$
for all $t\in (t_0,t_0+T).$ This is always possible since
$\mathfrak{L}(t)$ is of the order of distance between the two
points. Then the average value of $A$ over $\{x,z,t: x \in (x_0,
x_0 + \Delta x), t\in (t_0, t_0 + T), z = f(x,t)\}$ is defined as
\begin{eqnarray}\label{lineaverage}
\langle A \rangle_l = \frac{1}{\mathfrak{W}}\int\limits_{t_0}^{t_0
+ T}dt\int\limits_{x_0}^{x_0 + \Delta x}dl~A\,, \quad \mathfrak{W}
= \int\limits_{t_0}^{t_0 + T}dt~\mathfrak{L}(t) =
\int\limits_{t_0}^{t_0 + T}dt\int\limits_{x_0}^{x_0 + \Delta
x}dl\,,
\end{eqnarray}
\noindent $dl$ being the front line element, $dl = Ndx.$

Using the operation introduced, the quantities $\bm{\tau},\bm{n}$
defined on the front, as well as the flame front position itself,
can be decomposed into two parts corresponding to the scales $L_0$
and $\lambda_{\rm c},$ in a way analogous to
Eqs.~(\ref{01expansion})--(\ref{pexpansion}):
\begin{eqnarray}\label{ntfexpansion}&&
\bm{\tau} = \bm{\tau}_0 + \bm{\tau}_1\,, \quad
(n_x,n_z)=(-\tau_z,\tau_x)\,, \quad f = f_0 + f_1\,, \quad
\langle\bm{\tau}_1\rangle_l = 0\,, \quad \langle f_1\rangle_l =
0\,,
\\&& \bm{\tau}_{0,1} = \bm{\tau}_{0,1}^{(0)} + O(\varepsilon)\,,
\quad f_{0} = f_{0}^{(0)} + O(\varepsilon)\, \quad f_1 =
O(\varepsilon). \label{ntforders}
\end{eqnarray}
\noindent To obtain jump conditions for the quantities
$\bm{v}_0^{(0)},$ $p_0^{(0)},$ expansions
(\ref{01expansion})--(\ref{pexpansion}) and (\ref{ntfexpansion})
should be inserted into Eqs.~(\ref{conserv1})--(\ref{conserv3}),
with the subsequent averaging of the latter along the flame front.

Let us begin with the jump of the normal component of the gas
velocity. In view of Eq.~(\ref{cont}), one can introduce the
stream function $\psi = \psi(x,z,t)$ according to
\begin{eqnarray}\label{vstream}
u = \frac{\partial \psi}{\partial x}\,, \quad w = - \frac{\partial
\psi}{\partial z}\,.
\end{eqnarray}
\noindent Using the operation of the bulk averaging
(\ref{average}), the function $\psi$ can be decomposed as
\begin{eqnarray}\label{psidec}
\psi = \psi_0 + \psi_1\,, \quad \langle\psi_1\rangle = 0\,.
\end{eqnarray}
\noindent It follows from Eqs.~(\ref{orders0}) that
\begin{eqnarray}
\psi_0 = O(1)\,, \quad \psi_1 = O(\varepsilon)\,.
\end{eqnarray}
\noindent Substituting Eq.~(\ref{psidec}) into
Eqs.~(\ref{vstream}), and averaging gives
\begin{eqnarray}
u_0 = \left\langle\frac{\partial (\psi_0 + \psi_1)}{\partial
x}\right\rangle = \frac{\partial \psi_0}{\partial x} +
\frac{1}{L^2 T} \int\limits_{t_0}^{t_0 + T} dt
\left.\int\limits_{z_0}^{z_0 + L}dz~\psi_1\right|_{x_0}^{x_0 + L}
= \frac{\partial \psi_0}{\partial x} + O\left(\frac{L_{\rm f
}}{L}\right)\,, \nonumber
\end{eqnarray}
\noindent and analogous equation for $w_0.$ Thus, up to $o(1)$
terms, one has
\begin{eqnarray}\label{u0}
u_0 = \frac{\partial \psi_0}{\partial x}\,, \quad w_0 = -
\frac{\partial \psi_0}{\partial z}\,.
\end{eqnarray}
\noindent Inserting Eqs.~(\ref{vstream}) into
Eq.~(\ref{conserv1}), and using Eq.~(\ref{taun}), one finds
\begin{eqnarray}\label{vnav1}
\left\langle(\bm{v}_{\pm}\bm{n})\right\rangle_l &=&
\frac{1}{\mathfrak{W}}\int\limits_{t_0}^{t_0 +
T}dt\int\limits_{x_0}^{x_0 + \Delta x}dx~\left(\frac{\partial
\psi}{\partial x} + \frac{\partial\psi}{\partial z}\frac{\partial
f}{\partial x }\right)_{\pm} =
\frac{1}{\mathfrak{W}}\int\limits_{t_0}^{t_0 +
T}dt\int\limits_{x_0}^{x_0 + \Delta x}dx~\frac{d\psi_{\pm}}{dx}
\nonumber\\ &=& \frac{1}{\mathfrak{W}}\left.\int\limits_{t_0}^{t_0
+ T}dt~\psi_{0\pm}\right|_{x_0}^{x_0 + \Delta x} + o(1)\,.
\end{eqnarray}
\noindent Since the variation of $\psi_0$ over space distances
$\sim L$ and time intervals $\sim T$ is small, taking into account
Eqs.~(\ref{ntforders}), (\ref{u0}), and neglecting
$O(\varepsilon)$ terms, one can write
\begin{eqnarray}\label{vnav2}
\frac{1}{\mathfrak{W}}\left.\int\limits_{t_0}^{t_0 +
T}dt~\psi_{0\pm}\right|_{x_0}^{x_0 + \Delta x} &=&
\frac{T}{\mathfrak{W}}\left[\psi_{0\pm}(x_0 + \Delta x, f_0(x_0 +
\Delta x,t_0),t_0) - \psi_{0\pm}(x_0,f_0(x_0,t_0),t_0)\right] \nonumber\\
&=& \frac{T}{\mathfrak{W}}\left(\frac{\partial\psi_0}{\partial x}
+ \frac{\partial\psi_0}{\partial z}\frac{\partial f_0}{\partial x
}\right)_{\pm}\Delta x = \frac{T}{\mathfrak{W}} \left(u_{0\pm} -
w_{0\pm}\frac{\partial f_0}{\partial x}\right)\Delta x\,.
\end{eqnarray}
\noindent Substituting this into Eq.~(\ref{conserv1}) gives
\begin{eqnarray}\label{uw0}
\left[u_{0+} - u_{0-} - (w_{0+} - w_{0-})\frac{\partial
f_0}{\partial x}\right]\frac{T\Delta x}{\mathfrak{W}} = (\theta -
1) + o(1)\,.
\end{eqnarray}
\noindent Note that
\begin{eqnarray}\label{vnav3}
n_{x0} = \left\langle - \frac{\partial f/\partial
x}{N}\right\rangle_l = -
\frac{1}{\mathfrak{W}}\int\limits_{t_0}^{t_0 +
T}dt\int\limits_{x_0}^{x_0 + \Delta x}dx~\left(\frac{\partial f_0
}{\partial x} + \frac{\partial f_1 }{\partial x}\right) = -
\frac{T\Delta x}{\mathfrak{W}}\frac{\partial f_0 }{\partial x} +
o(1)\,,
\end{eqnarray}
\noindent and similarly,
\begin{eqnarray}\label{vnav4}
n_{z0} = \frac{T\Delta x}{\mathfrak{W}} + o(1)\,.
\end{eqnarray}
\noindent Hence, Eq.~(\ref{uw0}) can be rewritten as
\begin{eqnarray}
\left(\bm{v}_{0+}\bm{n}_0\right) -
\left(\bm{v}_{0-}\bm{n}_0\right) = (\theta - 1) + o(1)\,,
\nonumber
\end{eqnarray}
\noindent or, with the same accuracy,
\begin{eqnarray}\label{uw01}
\left(\bm{v}^{(0)}_{0+}\bm{n}^{(0)}_0\right) -
\left(\bm{v}^{(0)}_{0-}\bm{n}^{(0)}_0\right) = (\theta - 1) +
o(1)\,.
\end{eqnarray}
\noindent The inverse norm of $\bm{n_0}$
\begin{eqnarray}\label{veff}
\|\bm{n}_0\|^{-1} = \frac{
\displaystyle\frac{1}{T}\int\limits_{t_0}^{t_0 +
T}dt~\mathfrak{L}(t)} {\displaystyle\Delta x\sqrt{1 +
\left(\frac{\partial f_0}{\partial x}\right)^2}} \equiv
\mathfrak{U}
\end{eqnarray}
\noindent has a clear geometrical meaning. Namely, $(\mathfrak{U}
- 1)$ represents the relative increase of the flame front length
due to its small scale wrinkling.

Let us turn to the examination of the remaining jump conditions.
Because of the $\varepsilon$-terms in the right hand sides of
Eqs.~(\ref{conserv2}), (\ref{conserv3}), which involve highly
nonlinear combinations of the flow variables and the function
$f(x,t),$ there seems to be a very little hope that the large
scale parts of the flow variables eventually decouple from their
small scale parts describing flame cellular structure.
Nevertheless, they do, as will be shown presently.

According to Eqs.~(\ref{orders0})--(\ref{orders2}) and analogous
estimates for the space-time derivatives of $f(x,t),$
\begin{eqnarray}\label{ordert1}
\frac{\partial f}{\partial x} &=& O(1), \quad \frac{\partial
f}{\partial t} = O(1), \\ \quad \frac{\partial^2 f}{\partial t^2}
&=& O\left(\frac{1}{\varepsilon}\right), \quad\frac{\partial^2
f}{\partial x\partial t} = O\left(\frac{1}{\varepsilon}\right),
\quad\frac{\partial^2 f}{\partial x^2} =
O\left(\frac{1}{\varepsilon}\right)\,,\label{ordert2}
\end{eqnarray}
\noindent the $\varepsilon$-terms are $O(1).$ Let us show first
that the average value of the right hand side of
Eq.~(\ref{conserv2}) along the flame front is actually $o(1).$
Using the evolution equation (\ref{evolution}), expression in the
parentheses in the right hand side of Eq.~(\ref{conserv2}) can be
rewritten as follows:\footnotemark \footnotetext{It was mentioned
after Eq.~(\ref{evolution}), that the $\varepsilon$-terms in the
jump conditions represent the leading order terms of the
asymptotic expansion in powers of $1/c_1\,.$ In transforming these
terms, therefore, one can use the evolution equation with the
$\varepsilon$-term omitted.}
\begin{eqnarray}&&
\hat{D}w_{-} + f'\hat{D}u_{-} + \frac{1}{N}\hat{D}f' \nonumber\\&&
= \dot{w}_- + f'\dot{u}_- + \frac{\dot{f}'}{N} + w_-w_-' +
\frac{(f'w_-)'}{N} + \frac{\left(f'\right)^2 u_-'}{N} + f'w_-u_-'
+ \frac{N'}{N} \nonumber\\&& = \dot{w}_- + f'\dot{u}_- +
\frac{\dot{f}'}{N} + w_-w_-' + \frac{\left(u_- - N -
\dot{f}\right)'}{N} + \frac{N^2 - 1}{N}u_-' \nonumber\\&& +
u_-'\left(u_- - N - \dot{f}\right) + \frac{N'}{N} \nonumber\\&& =
\dot{w}_- + \frac{\partial (f'u_-)}{\partial t} - (u_- \dot{f})' +
\frac{\left(u_-^2 + w_-^2\right)'}{2}\,,
\end{eqnarray}
\noindent where the dot and the prime denote differentiation with
respect to $t$ and $x,$ respectively. Hence, taking into account
the estimates (\ref{orders0}), (\ref{ordert1}), one has

$$\left\langle\frac{1}{N}\left(\hat{D}w_{-} + f'\hat{D}u_{-} +
\frac{1}{N}\hat{D}f'\right)\right\rangle_l = $$
\begin{eqnarray}&&
= \frac{1}{\mathfrak{W}}\left(\int\limits_{x_0}^{x_0 + \Delta x}
dx~[w_- + f'u_-]_{t_0}^{t_0 + T} + \int\limits_{t_0}^{t_0 + T}
dt~\left[- u_-\dot{f} + \frac{u_-^2 + w_-^2}{2} \right]_{x_0}^{x_0
+ \Delta x}\right) \nonumber\\&& = O\left(\frac{T_0}{T}\right) +
O\left(\frac{L_0}{\mathfrak{L}}\right).
\end{eqnarray}
\noindent In view of the choice of $T,\mathfrak{L},$ the right
hand side of Eq.~(\ref{conserv2}) turns out to be $O(\varepsilon
L_0/\mathfrak{L}) = O(L_{\rm f}/\mathfrak{L}) = o(1).$ Thus,
averaging of Eq.~(\ref{conserv2}) gives
$$\left\langle(\bm{v}_+\bm{\tau}) - (\bm{v}_-\bm{\tau})\right\rangle_l
= o(1)\,.$$ To the leading order in $\varepsilon,$ this equation
can be written as
\begin{eqnarray}\label{taujumpaux}
\left\langle\left(\left[\bm{v}^{(0)}_{0+} -
\bm{v}^{(0)}_{0-}\right]\bm{\tau}^{(0)}\right)\right\rangle_l +
\left\langle\left(\left[\bm{v}^{(0)}_{1+} -
\bm{v}^{(0)}_{1-}\right]\bm{\tau}^{(0)}\right)\right\rangle_l =
o(1)\,, \quad \bm{\tau}^{(0)} = \bm{\tau}_0^{(0)} +
\bm{\tau}_1^{(0)}\,.
\end{eqnarray}\noindent
It was mentioned in the end of Sec.~\ref{eqdecoupling} that in the
approximation considered, the large scale flows do not affect the
flame cellular structure. In particular, the two directions along
the flame front, $\bm{\tau}$ and $- \bm{\tau},$ are left
equivalent. This implies that up to $O(\lambda_{\rm
c}/\mathfrak{L})$ terms, the value of $\langle(\bm{v}^{(0)}_{1+} -
\bm{v}^{(0)}_{1-})\bm{\tau}^{(0)}\rangle_l$ must be invariant
under the change $\bm{\tau}^{(0)} \to - \bm{\tau}^{(0)}.$ Hence,
$$\left\langle\left(\left[\bm{v}^{(0)}_{1+} -
\bm{v}^{(0)}_{1-}\right]\bm{\tau}^{(0)}\right)\right\rangle_l = -
\left\langle\left(\left[\bm{v}^{(0)}_{1+} -
\bm{v}^{(0)}_{1-}\right]\bm{\tau}^{(0)}\right)\right\rangle_l =
0\,.$$ Thus, taking into account definition of $\bm{\tau}_1,$ we
have from Eq.~(\ref{taujumpaux})
\begin{eqnarray}\label{taujump}
\left(\left[\bm{v}^{(0)}_{0+} -
\bm{v}^{(0)}_{0-}\right]\bm{\tau}_0^{(0)}\right) = o(1)\,.
\end{eqnarray}\noindent

Consider next the pressure jump, Eq.~(\ref{conserv3}). The last
term in the right hand side of this equation can be transformed
as\footnotemark\footnotetext{See the footnote 2.}
\begin{eqnarray}&&\label{pjumpaux}
\ddot{f} + 2w_-\dot{f}' + w_-^2 f'' + 2\hat{D}N - \frac{f'N'}{N}
\nonumber\\&& = \ddot{f} + 2w_-\dot{f}' + \left(w_-^2 f'\right)' -
2w_-'(u_- - N -\dot{f}) + 2\left(\dot{N} + w_-N' +
\frac{f'N'}{N}\right) - \frac{f'N'}{N} \nonumber\\&& = \ddot{f} +
2\dot{N} + 2(w_-\dot{f})' + 2(w_-N)' +\left(w_-^2 f'\right)' -
2w_-'u_- + \frac{f'N'}{N} \nonumber\\&& = \ddot{f} + 2\dot{N} -
(w_-^2 f')' + 2w_-u_-' + \frac{f'N'}{N}\,.
\end{eqnarray}
\noindent As before, the first three terms give rise to $o(1)$
terms upon averaging. The remaining two terms, however, do not
reduce to the full $x$- or $t$-derivatives. Their contribution is,
therefore, $O(1),$ in general. Notice that the fourth term is
quadratic in the gas velocity. This fact can be used to show that
its contribution is independent of the functions $\bm{v}_{0-},
p_{0-}.$ Indeed, one has, to the leading order in $\varepsilon,$
\begin{eqnarray}
\varepsilon\left\langle\frac{w_- u_-'}{N}\right\rangle_l &=&
\frac{\varepsilon}{\mathfrak{W}}\int\limits_{t_0}^{t_0 +
T}dt\int\limits_{x_0}^{x_0 + \Delta x} dx~\left(w_{0-}^{(0)} +
w_{1-}^{(0)}\right) \left(u_{0-}^{(0)} + u_{1-}^{(0)}\right)' \nonumber\\
&=& \frac{\varepsilon}{\mathfrak{W}}\int\limits_{t_0}^{t_0 +
T}dt\int\limits_{x_0}^{x_0 + \Delta x} dx~\left(w_{0-}^{(0)} +
w_{1-}^{(0)}\right)u_{1-}^{(0)\prime} + O(\varepsilon)
\nonumber\\&=&
\frac{\varepsilon}{\mathfrak{W}}\int\limits_{t_0}^{t_0 + T}
dt~\left[w_{0-}^{(0)}u_{1-}^{(0)}\right]_{x_0}^{x_0 + \Delta x} +
\frac{\varepsilon}{\mathfrak{W}}\int\limits_{t_0}^{t_0 +
T}dt\int\limits_{x_0}^{x_0 + \Delta x}
dx~w_{1-}^{(0)}u_{1-}^{(0)\prime} + O(\varepsilon)\nonumber\\&=&
\frac{\varepsilon}{\mathfrak{W}}\int\limits_{t_0}^{t_0 +
T}dt\int\limits_{x_0}^{x_0 + \Delta x}
dx~w_{1-}^{(0)}u_{1-}^{(0)\prime} + O\left(\frac{L_{\rm
f}}{\mathfrak{L}}\right)\,. \nonumber
\end{eqnarray}
\noindent Thus, up to $o(1)$ terms, the average value of
$\varepsilon w_- u_-'/N$ turns out to be independent of the
functions $\bm{v}_0^{(0)}.$ The quantities $w_{1-}^{(0)},$
$u_{1-}^{(0)}$ describe variations of the fuel velocity along the
front cell in zero order approximation with respect to $L_{\rm
f}/L_0,$ {\it i.e.,} when the influence of the large scale flow on
the local flame structure is completely neglected. In particular,
the value of $\langle \varepsilon w_- u_-'/N\rangle_l$ is
independent of the coordinate $x_0$ as well as of the time instant
$t_0.$ Denote this constant by $\alpha_1/2\,.$\footnotemark
\begin{footnotetext}{
At first sight, $\langle \varepsilon w_- u_-'/N\rangle_l$ depends
on the choice of orientation of the coordinate axes, while the
scalar pressure jump must be independent of this choice. It is
easy to see, however, that within the accuracy of the above
calculations, $\langle \varepsilon w_- u_-'/N\rangle_l$ is
actually coordinate-invariant. In fact, under rotations of the
coordinate system, $w_-,$ $u_-$ transform as
\begin{eqnarray}
w_- &\to& \cos\varphi~w_- + \sin\varphi~u_-\,, \nonumber\\ u_-
&\to& -\sin\varphi~w_- + \cos\varphi~u_-\,, \nonumber
\end{eqnarray}
where $\varphi$ is the rotation angle. This transformation leaves
$\langle \varepsilon w_- u_-'/N\rangle_l$ unchanged:
\begin{eqnarray}
\langle \varepsilon w_- u_-'/N\rangle_l &\to&
\frac{\varepsilon}{\mathfrak{W}}\int dt\int~(\cos\varphi~w_- +
\sin\varphi~u_-)d( - \sin\varphi~w_- + \cos\varphi~u_-) \nonumber\\
&=& \langle \varepsilon w_- u_-'/N\rangle_l + O\left(\frac{L_{\rm
f}}{\mathfrak{L}}\right). \nonumber
\end{eqnarray}\noindent
Definition of $\alpha_1$ can be written also in an explicitly
invariant form:
$$\alpha_1 = \varepsilon\left\langle\frac{w_-u_-' - u_-w_-'}{N}\right\rangle_l.\,$$
}\end{footnotetext}

It remains to find the contribution of the last term in
Eq.~(\ref{pjumpaux}). We have
\begin{eqnarray}
\varepsilon\left\langle\frac{f'N'}{N^2}\right\rangle_l =
\frac{\varepsilon}{\mathfrak{W}}\int\limits_{t_0}^{t_0 +
T}dt\int\limits_{x_0}^{x_0 + \Delta x} dx~\frac{f'N'}{N} = -
\varepsilon\left\langle\left(\frac{f'}{N}\right)'\right\rangle_l +
O\left(\frac{L_{\rm f}}{\mathfrak{L}}\right).
\end{eqnarray}
The quantity $\left(f'/N\right)'$ is nothing but the flame front
curvature $k,$ $$k = \frac{f''}{N^3}\,.$$ Using the definitions
(\ref{taun}), (\ref{ntfexpansion}), one can write $$\langle
k\rangle_l = \left\langle\left(\frac{f'}{N}\right)'\right\rangle_l
= \langle\tau_z'\rangle_l = \langle\tau_{z0}'\rangle_l +
\langle\tau_{z1}'\rangle_l = \tau_{z0}' +
\langle\tau_{z1}'\rangle_l\,.$$ According to Eqs.~(\ref{ordert1}),
(\ref{ordert2}), $\tau_{z0}' = O(1),$ $\tau_{z1}' =
O(1/\varepsilon).$ Thus, to the leading order in $\varepsilon,$
$$\varepsilon\langle k\rangle_l = \varepsilon
\left\langle\tau_{z1}^{(0)\prime}\right\rangle_l = O(1)\,.$$
Similarly to $\bm{v}_1^{(0)},$ $\tau_{z1}^{(0)}$ describes
geometry of a front cell neglecting the influence of the large
scale flame structure on it. Hence, the value of
$\langle\tau_{z1}^{(0)\prime}\rangle_l$ is independent of the
particular choice of the point $x_0$ on the flame front (and of
the time instant $t_0$). Denoting this constant by $k_1,$ we thus
obtain from Eq.~(\ref{conserv3}) the following expression for the
jump of $p_0^{(0)}$ at the flame front
\begin{eqnarray}\label{presjump}
p_{0+}^{(0)} - p_{0-}^{(0)} = - \frac{\theta - 1}{\theta}G
f^{(0)}_0 - (\theta - 1) - \pi_1 + \varepsilon [(\theta - 1) -
\ln\theta] k_1 + \varepsilon\ln\theta\alpha_1 + o(1)\,,
\end{eqnarray}\noindent where $\pi_1$ is another constant defined by
$$\pi_1 = \left\langle p_{1+}^{(0)} - p_{1-}^{(0)}\right\rangle_l\,.$$ It is
independent of $x_0, t_0$ on the same grounds as $\alpha_1, k_1.$

Finally, we have to average the evolution equation
(\ref{evolution}). As before, contribution of the $\varepsilon$
term on the right hand side of this equation is $o(1).$
Furthermore,
$$\left\langle\frac{1}{N}\frac{\partial f}{\partial t
}\right\rangle_l =
\frac{1}{\mathfrak{W}}\left.\int\limits_{x_0}^{x_0 + \Delta
x}dx~f_0\right|_{t_0}^{t_0 + T} + o(1) = \frac{T\Delta x
}{\mathfrak{W}}\frac{\partial f_0}{\partial t} + o(1)\,.$$ Using
Eqs.~(\ref{vnav1})--(\ref{vnav4}), we find
\begin{eqnarray}\label{avevolution}
(\bm{v}_{0-}\bm{n}_0) - n_{z0}\dot{f}_0 = 1 + o(1)\,.
\end{eqnarray}\noindent

Equations (\ref{uw01}), (\ref{taujump}), (\ref{presjump}), and
(\ref{avevolution}) constitute the proof of the second part of the
decoupling theorem. To make this more transparent, let us rewrite
these equations in a more convenient notation. The jump conditions
for the large scale parts of the flow variables read\footnotemark
\footnotetext{The vector quantities denoted by Gothic letters are
designated with arrows.}
\begin{eqnarray}&&
\left(\vec{\mathfrak{v}}_{+}\vec{\mathfrak{n}}\right)
-\left(\vec{\mathfrak{v}}_{-}\vec{\mathfrak{n}}\right) = (\theta -
1)\mathfrak{U} + o(1)\,,\label{finalj1}\\&&
\left(\vec{\mathfrak{v}}_{+}\vec{\mathfrak{t}}\right) -
\left(\vec{\mathfrak{v}}_{-}\vec{\mathfrak{t}}\right) =
o(1)\,,\label{finalj2}\\&& \mathfrak{p}_+ - \mathfrak{p}_- = -
\frac{\theta - 1}{\theta}G \mathfrak{f} + \Pi +
o(1)\,,\label{finalj3}
\end{eqnarray}\noindent
where
\begin{eqnarray}
\vec{\mathfrak{n}} &=& \frac{\bm{n}_0^{(0)}}{\|\bm{n}_0^{(0)}\|} =
\left(- \frac{\mathfrak{f}'}{\EuScript{N}}\,,
\frac{1}{\EuScript{N}}\right)\,, \quad\mathfrak{f} = f^{(0)}_0\,,
\quad \EuScript{N} = \sqrt{1 + \left(\mathfrak{f}'\right)^2}\,,
\quad \vec{\mathfrak{v}} = \bm{v}_0^{(0)}\,,\nonumber\\
\quad\vec{\mathfrak{t}} &=&
\frac{\bm{\tau}_0^{(0)}}{\|\bm{\tau}_0^{(0)}\|} = \left(
\frac{1}{\EuScript{N}}\,,
\frac{\mathfrak{f}'}{\EuScript{N}}\right)\,, \quad
\|\bm{n}_0^{(0)}\| = \|\bm{\tau}_0^{(0)}\| = \mathfrak{U}^{-1}\,,\nonumber\\
\quad \mathfrak{p} &=& p_0^{(0)}\,, \quad\Pi = - (\theta - 1) -
\pi_1 + \varepsilon [(\theta - 1) - \ln\theta] k_1 +
\varepsilon\ln\theta\alpha_1\,.\nonumber
\end{eqnarray}\noindent
The evolution equation takes the form
\begin{eqnarray}\label{gotevolution}
\left(\vec{\mathfrak{v}}_{-}\vec{\mathfrak{n}}\right) -
\frac{\dot{\mathfrak{f}}}{\EuScript{N}} = \mathfrak{U} + o(1)\,.
\end{eqnarray}\noindent
As we have seen, in zero order approximation with respect to
$\varepsilon,$ the quantities $k_1,$ $\alpha_1,$ $\pi_1,$ and
$\mathfrak{U}$ entering these equations are independent of the
coordinate $x_0$ and the time instant $t_0.$ These constants can
in principle be calculated provided that the exact small scale
flame structure is known. However, as it follows from the above
equations, this information is actually unnecessary, because
dynamics of the large scale fields are independent of the specific
values of these constants. Indeed, since the gas pressure enters
dynamical equation only through its gradient, it is determined up
to a constant. Therefore, the constants $k_1,$ $\alpha_1,$
$\pi_1,$ are irrelevant. Unlike these, however, the constant
$\mathfrak{U}$ has a direct physical meaning. As is seen from
Eq.~(\ref{gotevolution}), $\mathfrak{U}$ plays the role of the
effective dimensionless velocity of the curved flame propagation.
According to our choice of units, the gas velocity is scaled on
the adiabatic velocity of a plane flame front $U_{\rm f}$ which is
also used to define the units of pressure, time, and length [see
the definition (\ref{l0}) of $L_0$]. In analyzing the large scale
flame dynamics, it is more natural to choose $\mathfrak{U}\cdot
U_{\rm f},$ rather than $U_{\rm f},$ as the velocity unit. Then
$\mathfrak{U}$ disappears from the jump conditions at the flame
front, as well as from the flow equations (\ref{flow1cont}),
(\ref{aveuler1}) in the bulk, which thus take the form of the
equations governing propagation of a zero-thickness flame in an
ideal fluid at constant speed $\mathfrak{U}U_{\rm f}$ with respect
to the fuel, $\mathfrak{f}$ being the flame front position, while
$\vec{\mathfrak{n}}$ and $\vec{\mathfrak{t}}$ the normal and
tangential unit vectors to $\mathfrak{f},$ respectively. The
decoupling theorem is proved.

\section{Nonlinear flame stabilization in gravitational
field}\label{example}

The proved theorem considerably widens the scope of issues in
flame dynamics accessible for analytical investigation. One of the
most important consequences of the decoupling theorem is that
unlike the local cellular dynamics, the large scale flame dynamics
can be investigated in the framework of perturbation expansion
with respect to the flame front slope, provided that the external
field exerts a stabilizing influence on the flame. This fact will
be illustrated below in the case of a flame propagating in an
initially quiescent fluid in the direction of the gravitational
field. As is well known (see, {\it e.g.}, Ref.~\cite{zel}), in
this case gravity plays the stabilizing role at the linear stage
of development of the LD-instability. Our aim will be to determine
the role of the nonlinear effects, and to explore the possibility
of a full stabilization of the curved flame front by the
gravitational field.

We will follow the general method of deriving weakly nonlinear
equations for the flame front position, developed in
Ref.~\cite{kazakov3}. It consists in bringing the system of
hydrodynamic equations together with the jump conditions at the
flame front to the so-called transverse representation in which
dependence of all flow variables on the coordinate in the
direction of flame propagation ($z$) is rendered purely
parametric, and then reducing this system to a single equation for
the front position. The calculation in the presence of gravity is
very similar to that in the case of a freely propagating flame.
Therefore, derivation of the equation will be only sketched below,
referring the reader to the work \cite{kazakov3} for more detail.

As was mentioned in the end of the preceding section, it is
natural to take $\mathfrak{U}U_{\rm f}$ as the velocity unit,
redefining accordingly the units of length, time, and pressure to
$$\frac{\left(\mathfrak{U}U_{\rm f}\right)^2}
{\|\vec{g}\|} \equiv \mathfrak{L}_0\,,
\quad\frac{\mathfrak{L}_0}{\mathfrak{U}U_{\rm f}}\,, \quad
\left(\rho_{\rm u}\mathfrak{U}U_{\rm f}\right)^2\,,$$
respectively. For simplicity, designation of the flow variables as
well as space coordinates and time will be left unchanged. Then
the bulk equations read\footnotemark \footnotetext{For brevity,
the $o(1)$ symbols will be omitted in what follows.}
\begin{eqnarray}\label{contgot}
{\rm div}\mathfrak{v} &=& 0\,, \\
\dot{\vec{\mathfrak{v}}} + \left(\vec{\mathfrak{v}}
\vec\nabla\right)\vec{\mathfrak{v}} &=& -
\frac{\vec\nabla\mathfrak{p}}{\varrho} + \vec{\mathfrak{G}}\,,
\quad \vec{\mathfrak{G}} =
\frac{\bm{g}\mathfrak{L}_0}{\mathfrak{U}^2}\,, \quad
\|\vec{\mathfrak{G}}\| = 1\,.\label{aveulergot}
\end{eqnarray} \noindent
Since the fuel is assumed initially quiescent, the flow is
potential upstream, and the general solution of
Eqs.~(\ref{contgot}), (\ref{aveulergot}) can be readily written
down. In the reference frame of an initially plane flame front,
\begin{eqnarray}\label{solup1}
\mathfrak{u} &=& 1 + \int\limits_{- \infty}^{+ \infty}d k
~\mathfrak{u}_k \exp (|k|z + i k x)\,, \\ \label{solup2}
\mathfrak{w} &=& \hat{\rm H}(\mathfrak{u} - 1)\,, \\
\label{solup3} \dot{\mathfrak{u}} &+& \hat{\Phi}\left(\mathfrak{p}
+ \mathfrak{G}z\right) + \frac{\hat{\Phi}}{2}(\mathfrak{u}^2 +
\mathfrak{w}^2) = 0\,, \\
\quad \mathfrak{w}&\equiv&\mathfrak{v}_x\,, \quad
\mathfrak{u}\equiv\mathfrak{v}_z\,, \quad\mathfrak{G} \equiv -
\mathfrak{G}_z\,.\nonumber
\end{eqnarray}
\noindent Here $\hat{\rm H}$ denotes the Hilbert operator defined
as
\begin{eqnarray}\label{hilbert}
\hat{\rm H}\exp(i k x) &=& i~{\rm sign}(k)\exp(i k x)\,, \quad
k\ne 0\,,\\
{\rm sign}(k) &\equiv& \frac{k}{|k|}\,. \nonumber
\end{eqnarray}
The LD-operator $\hat{\Phi}$ is related to $\hat{\rm H}$ by
$\hat{\Phi} = - \hat{\rm H}\cdot\partial/\partial x\,.$ Equation
(\ref{solup3}) is nothing but the Bernoulli equation written in
the transverse form.

Because of the vorticity produced by the curved flame front, the
flow of products of combustion is not potential. Nevertheless, the
following transverse relation between the flow variables
downstream can be obtained from Eqs.~(\ref{contgot}),
(\ref{aveulergot}) at the second order of nonlinearity
\begin{eqnarray}\label{streamgot}&&
\dot{\mathfrak{u}} - \theta\mathfrak{w}' -
\hat{\Phi}\left(\theta\mathfrak{p} + \frac{(\mathfrak{u} -
\theta)^2 + \mathfrak{w}^2}{2}\right) +
\mathfrak{w}\left(\mathfrak{u}' +
\frac{\dot{\mathfrak{w}}}{\theta} + \mathfrak{p}'\right) = 0\,,
\end{eqnarray}
\noindent Equations (\ref{finalj1})--(\ref{gotevolution}),
(\ref{solup2}), (\ref{solup3}), and (\ref{streamgot}) constitute
the closed system of equations describing flame dynamics in the
transverse representation. It can be reduced to the following
equation for the function $\mathfrak{f}$
\begin{eqnarray}&&\label{nonstationary}
\hspace{-1cm} (\theta + 1)\ddot{\mathfrak{f}} +
2\theta\hat{\Phi}\dot{\mathfrak{f}} + \theta (\theta -
1)\mathfrak{f}'' + (\theta - 1)\mathfrak{G}\hat{\Phi}\mathfrak{f}
+ \left(\theta - \frac{(\theta -
1)^2}{2}\right)\hat{\Phi}(\mathfrak{f}')^2 \nonumber\\&&
\hspace{-1cm} + \frac{(\theta -
1)^2}{\theta}\mathfrak{G}(\mathfrak{f}')^2 + \left(\theta +
\frac{1}{\theta}\right)\left(\mathfrak{f}'\dot{\mathfrak{f}}' +
\dot{\mathfrak{f}}'\hat{\rm H}\dot{\mathfrak{f}}\right) +
\frac{\theta - 1}{2}\hat{\Phi}\left(\dot{\mathfrak{f}}^2 +
\left(\hat{\rm H}\dot{\mathfrak{f}}\right)^2\right) \nonumber\\&&
\hspace{-1cm} + (3\theta - 1)\hat{\Phi}\left(\mathfrak{f}'\hat{\rm
H}\dot{\mathfrak{f}}\right) + \left(2\theta - 1 +
\frac{1}{\theta}\right)\mathfrak{f}'\hat{\rm H}\ddot{\mathfrak{f}}
- \frac{\theta - 1}{\theta}\left(\hat{\rm H}\ddot{\mathfrak{f}} +
\mathfrak{G}\mathfrak{f}'\right)\hat{\rm H}\dot{\mathfrak{f}} =
0\,.
\end{eqnarray}
\noindent The linear terms in this equation reproduce the
well-known equation
$$(\theta + 1)\ddot{\mathfrak{f}} +
2\theta\hat{\Phi}\dot{\mathfrak{f}} + \theta (\theta -
1)\mathfrak{f}'' + (\theta - 1)\mathfrak{G}\hat{\Phi}\mathfrak{f}
= 0\,,$$ from which it follows that at the linear stage of
development of the LD-instability, the gravitational field plays
the stabilizing role in the case of flame propagation in the
direction of $\vec{\mathfrak{G}}$ ($\mathfrak{G} = + 1$), and
destabilizing in the opposite case ($\mathfrak{G} = -1$). To
determine the role of the nonlinear terms, let us assume that
there exists a stationary regime of flame propagation. It should
be stressed that this assumption concerns only the the large scale
front structure described by the function $\mathfrak{f}.$ The
local cellular structure does not need to be stationary. Then
Eq.~(\ref{nonstationary}) simplifies to
\begin{eqnarray}\label{stationary}&&
\theta (\theta - 1)\mathfrak{f}'' + (\theta -
1)\mathfrak{G}\hat{\Phi}\mathfrak{f} + \left(\theta -
\frac{(\theta - 1)^2}{2}\right)\hat{\Phi}(\mathfrak{f}')^2 +
\frac{(\theta - 1)^2}{\theta}\mathfrak{G}(\mathfrak{f}')^2 = 0\,.
\end{eqnarray}
\noindent It is not difficult to see that the nonlinear term
proportional to $\mathfrak{G}$ exerts a stabilizing influence on
the flame if $\mathfrak{G} = + 1.$ Indeed, if we take
\begin{eqnarray}\label{lanzats}&&
\mathfrak{f}(x,t) \sim e^{\sigma t}\sin(k x),
\end{eqnarray}
\noindent Eq.~(\ref{nonstationary}) can be roughly considered as a
``dispersion relation'' for the increment $\sigma.$ As we see, the
nonlinear term decreases $\sigma$ if $\mathfrak{G} = + 1,$ and
vice versa. Therefore, Eq.~(\ref{stationary}) can only have
solutions if $\mathfrak{G} = + 1,$ and the question of whether the
flame can be stabilized by the gravitational field reduces to the
question of existence of nontrivial solutions to this
equation.\footnotemark \footnotetext{The last term in the
non-stationary equation (\ref{nonstationary}) is also proportional
to $\mathfrak{G}.$ Unlike the other two, it has a destabilizing
effect on the flame front in the case $\mathfrak{G} = + 1. $
Indeed, using the definition of the Hilbert operator
(\ref{hilbert}), and substituting expression (\ref{lanzats}) into
this term, one finds
$$ - \frac{\theta - 1}{\theta}\mathfrak{f}'
\hat{\rm H}\dot{\mathfrak{f}} = - \frac{\theta - 1}{\theta}\sigma
|k| \cos^2(k x) < 0.$$ This term, however, is irrelevant to the
issue of existence of stationary configurations, since it
contains~$\dot{\mathfrak{f}}.$ Whether the gravitational field has
an overall stabilizing effect, or not, depends on solvability of
the stationary equation (\ref{stationary}).}

Equation (\ref{stationary}) is a nonlinear integro-differential
equation with respect to $\phi = \mathfrak{\mathfrak{f}}'.$ To
solve this equation, we first transform it as follows. Let us
rewrite the nonlinear term
$\hat{\Phi}(\mathfrak{\mathfrak{f}}')^2$ iterating
Eq.~(\ref{stationary}) with respect to
$\mathfrak{\mathfrak{f}}'',$ {\it i.e.,} substituting
$$\mathfrak{f}'' = - \frac{\mathfrak{G}}{\theta}\hat{\Phi}\mathfrak{f} + O(\mathfrak{f}'^2).$$
One has
$$\hat{\Phi}(\mathfrak{f}')^2 \equiv - \hat{\rm H}\frac{\partial(\mathfrak{f}')^2}{\partial\eta}
= - 2\hat{\rm H}(\mathfrak{f}'\mathfrak{f}'') =
\frac{2\mathfrak{G}}{\theta}\hat{\rm
H}(\mathfrak{f}'\hat{\Phi}\mathfrak{f}) + O(\mathfrak{f}'^3)\,.$$
Using the well-known identity
\begin{eqnarray} 2\hat{\rm
H}\{\psi\hat{\rm H}\psi\} = (\hat{\rm H}\psi)^2 - \psi^2\,,
\nonumber
\end{eqnarray}
\noindent we find
$$\hat{\Phi}(\mathfrak{f}')^2 = - \frac{\mathfrak{G}}{\theta} \{(\hat{\rm H}\mathfrak{f}')^2 - (\mathfrak{f}')^2\}
+ O(\mathfrak{f}'^3)\,.$$  Hence, within the accuracy of the
second order, Eq.~(\ref{stationary}) takes the form
\begin{eqnarray}\label{stationary1}&&
\mathfrak{f}'' + \alpha\hat{\Phi}\mathfrak{f} +
\beta(\mathfrak{f}')^2 -
\gamma\left(\hat{\Phi}\mathfrak{f}\right)^2 = 0\,,
\end{eqnarray}
\noindent where
$$\alpha = \frac{\mathfrak{G}}{\theta}\,, ~~\beta =
\frac{\mathfrak{G}}{\theta^2(\theta -1)} \left(\theta +
\frac{(\theta - 1)^2}{2}\right)\,, ~~\gamma =
\frac{\mathfrak{G}}{\theta^2(\theta -1)} \left(\theta -
\frac{(\theta - 1)^2}{2}\right)\,.$$ \noindent In connection with
the transformation performed, it is worth to emphasize validity of
the weak nonlinearity expansion when applied to the investigation
of the large scale flame dynamics. As was shown in detail in
Refs.~\cite{kazakov1,kazakov2}, this expansion turns out to be
self-contradictory in the case of a freely propagating stationary
flame treated in the framework of the thin front model. This fact
can be seen directly from Eq.~(\ref{stationary}) with
$\mathfrak{G} = 0,$ in which case this equation reduces to the
equality of two quantities of apparently different orders --
$\mathfrak{f}''$ and
$\hat{\Phi}(\mathfrak{f}')^2\,.$\footnotemark\footnotetext{Perhaps,
it is worth to stress once more that despite similarity of
Eq.~(\ref{nonstationary}) with $\mathfrak{G} = 0$ to that obtained
in Ref.~\cite{kazakov3}, its meaning is completely different. In
the notation of Sec.~\ref{jdecoupling}, the latter equation
determines the function $f_1^{(0)},$ while
Eq.~(\ref{nonstationary}) -- the function $f_0^{(0)}.$} Only if
$\theta \to 1$ does the weak nonlinearity expansion of stationary
flames make sense, since then $\mathfrak{f}' = O(\theta - 1),$
$\mathfrak{f}'' = O((\theta - 1)^2),$ so both terms in
Eq.~(\ref{stationary}) are $O((\theta - 1)^3)$ quantities. On the
contrary, in the presence of the gravitational field,
$\mathfrak{f}'$ {\it can} be treated as the first order quantity
when $(\theta - 1)$ is not small, because $\theta\mathfrak{f}''$
and $\mathfrak{G}\hat{\Phi}\mathfrak{f}$ are of the first order in
this case, and nothing prevents their sum from being formally a
second order quantity. In fact, $(\theta - 1)$ {\it must} be
finite in the latter case, since the second order term $(\theta -
1) \mathfrak{G}\hat{\Phi}\mathfrak{f}$ would be the only second
order term in Eq.~(\ref{stationary}) otherwise. As we will see
below, solutions to this equation turn out to be unbounded for
$\theta \to 1.$

Turning back to Eq.~(\ref{stationary1}), let us show first of all
that its solutions, if any, are non-periodic. Notice that if
$\mathfrak{f}(x)$ is a periodic function, then $[\hat{\rm
H}(\mathfrak{f} - \bar{\mathfrak{f}})](x),$ where
$\bar{\mathfrak{f}}$ is the mean value of $\mathfrak{f},$ is also
periodic with the same period [substraction of
$\bar{\mathfrak{f}}$ is necessary, because $\hat{\rm H}$ is
undefined on constants, see Eq.~(\ref{hilbert})]. Let us integrate
Eq.~(\ref{stationary1}) over period. The first two terms in this
equation give rise to zero:
\begin{eqnarray}
\int dx~\mathfrak{f}''&=& \left. \mathfrak{f}'\right| = 0\,,
\nonumber\\
\int dx~\hat{\Phi}\mathfrak{f} &=& - \int dx~\hat{\rm
H}\mathfrak{f}' = - \int dx~\hat{\rm H}(\mathfrak{f} -
\bar{\mathfrak{f}})'
\nonumber\\
&=& - \int dx~\left\{\hat{\rm H}(\mathfrak{f} -
\bar{\mathfrak{f}})\right\}' = \left.\hat{\rm H}(\mathfrak{f} -
\bar{\mathfrak{f}})\right| = 0\,. \nonumber
\end{eqnarray}
\noindent On the other hand, using unitarity of the Hilbert
operator, we find
\begin{eqnarray}
\int dx \left\{\beta(\mathfrak{f}')^2 - \gamma\left(\hat{\rm
H}\mathfrak{f}'\right)^2\right\} = \int dx~(\beta -
\gamma)(\mathfrak{f}')^2 > 0\,. \nonumber
\end{eqnarray}
\noindent Of course, the absence of periodic solutions could be
inferred already from Eq.~(\ref{stationary}). It is seen that such
solutions are forbidden by the positive definite nonlinear term
proportional to $\mathfrak{G}.$

Non-periodic solutions can be found in the form of the pole
decomposition for the function $\phi=\mathfrak{f}'$
\begin{eqnarray}\label{poledec}
\phi = a~\sum\limits_{k=1}^{2P}\frac{1}{x - x_k}\,,
\end{eqnarray}
\noindent where the value of the amplitude $a,$ as well as
position of $P$ pares of the complex conjugate poles $x_k,$ are to
be determined by substituting this decomposition into
Eq.~(\ref{stationary1}). Using the definition of Hilbert operator,
one can show that
\begin{eqnarray}
\hat{\rm H}\phi &=& -i~a~\sum\limits_{k=1}^{2P}\frac{{\rm
sign}({\rm Im}~x_k)} {x - x_k}\,. \nonumber
\end{eqnarray}
\noindent It is not difficult to verify that
Eq.~(\ref{stationary1}) is satisfied by (\ref{poledec}) provided
that
$$a = \frac{1}{\beta + \gamma}\,,$$ and $x_k$ satisfy the
following system of algebraic equations
\begin{eqnarray}\label{poles}
i\alpha(\beta + \gamma) +
2\sum\limits_{\genfrac{}{}{0pt}{}{l=1}{l\ne k}}^{2P}
\frac{\beta~{\rm sign}({\rm Im}~x_k) + \gamma~{\rm sign}({\rm
Im}~x_l)}{x_k - x_l} = 0\,, \qquad k = 1,...,2 P\,.
\end{eqnarray}
\noindent Evidently, this system is only consistent if
$\mathfrak{G} = + 1.$ Indeed, in the case of a pare of complex
conjugate poles $x_1,$ $x_2 = x_1^{*},$ one has, assuming ${\rm
Im}~x_1>0,$
$${\rm Im}~x_1 = \frac{\beta - \gamma}{\alpha(\beta + \gamma)}
= \frac{(\theta - 1)^2}{2\mathfrak{G}}\,,$$ which is consistent
with the assumed positivity of ${\rm Im}~x_1$ if $\mathfrak{G} = +
1.$ It is not difficult to show that the same is true in the
general case of arbitrary number of poles. Consider the imaginary
part of Eq.~(\ref{poles}) corresponding to the pole uppermost in
the complex plane, and take into account that $\beta
> \gamma.$ The fact that the system (\ref{poles}) is inconsistent
for $\mathfrak{G} = - 1$ does not mean, of course, that the
nonlinear stabilization is impossible in this case. Investigation
of such a possibility requires account of higher order
corrections.

In the case of $P=1,$ one has, furthermore,
$$\mathfrak{f}' = \frac{1}{\beta + \gamma}\left(\frac{1}{x - x_1}
+ \frac{1}{x - x_1^*}\right) = \frac{2}{\beta + \gamma} \frac{(x -
{\rm Re}~x_1)}{(x - {\rm Re}~x_1)^2 + ({\rm Im}~x_1)^2}\,,$$ and
therefore,
\begin{eqnarray}
\mathfrak{f} = \frac{\theta(\theta - 1)}{2}\ln\left\{ (x - x_0)^2
+ \frac{(\theta - 1)^4}{4} \right\}\,, \qquad x_0 \equiv {\rm
Re}~x_1\,. \nonumber
\end{eqnarray}
\noindent The two-pole solutions for the cases $\theta = 5, 10$
and $x_0 = 0$ are shown in Fig.~2.

\begin{figure}
\scalebox{0.7}{\includegraphics{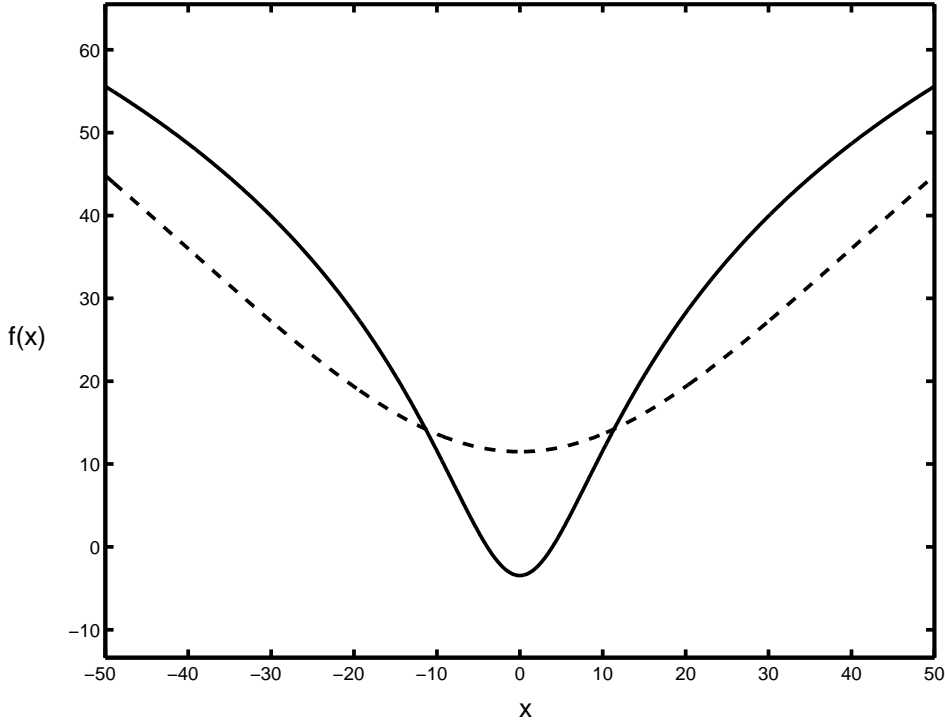}} \label{fig2}
\caption{Two-pole solutions of Eq.~(\ref{stationary1}) for $\theta
= 5$ (full line), and $\theta = 10$ (dashed line). Flames
propagate downwards.}
\end{figure}

Equation (\ref{stationary1}) is derived in the scope of the power
expansion with respect to the flame front slope $\phi.$ In the
case of the two-pole solution, $|\phi|$ takes its maximal value
$$\phi_{\rm m} = \frac{\theta}{\theta - 1}$$ at the points
$x_{\rm m} = x_0\pm {\rm Im}~x_1.$ We see that the developed weak
nonlinearity expansion is valid if $\theta$ is not too close to
unity. For realistic values of the expansion coefficient ($\theta
= 5-10$) $\phi_{\rm m} \approx 1.$

Next, consider the four-pole solution ($P = 2$). It has the form
\begin{eqnarray}
\mathfrak{f} = \frac{\theta(\theta - 1)}{2}\ln\left\{ \left[(x -
{\rm Re}~x_1)^2 + \left({\rm Im}~x_1\right)^2\right] \left[(x -
{\rm Re}~x_2)^2 + \left({\rm Im}~x_2\right)^2\right] \right\}\,.
\nonumber
\end{eqnarray}
\noindent Assuming that ${\rm Im}~x_{1,2} > 0,$ one has the
following equations for the position of poles $x_1,$ $x_2,$ $x_3 =
x_1^*,$ $x_4 = x_2^*$
\begin{eqnarray}
i~\alpha(\beta + \gamma) + 2\left\{\frac{\beta + \gamma}{x_1 -
x_2} + (\beta - \gamma)\left(\frac{1}{2i~{\rm Im}~x_1} +
\frac{1}{x_1 - x_2^*}\right)\right\} &=& 0\,,
\nonumber\\
i~\alpha(\beta + \gamma) + 2\left\{\frac{\beta + \gamma}{x_2 -
x_1} + (\beta - \gamma)\left(\frac{1}{2i~{\rm Im}~x_2} +
\frac{1}{x_2 - x_1^*}\right)\right\} &=& 0\,. \nonumber
\end{eqnarray}
\noindent Separating the real and imaginary parts, and rearranging
yields three equations for the four quantities ${\rm
Re}~x_{1,2}\,,$ ${\rm Im}~x_{1,2}$
\begin{eqnarray}
2\alpha\frac{\beta + \gamma}{\beta - \gamma} -
\left\{\frac{1}{{\rm Im}~x_1} + \frac{1}{{\rm Im}~x_2} + 4
\frac{{\rm Im}(x_1 + x_2)}{|x_1 - x_2^*|^2}\right\} &=& 0\,,
\label{4poles1}
\\
4~\frac{\beta + \gamma}{\beta - \gamma} \frac{{\rm Im}~(x_2 -
x_1)}{|x_1 - x_2|^2} + \frac{1}{{\rm Im}~x_2} - \frac{1}{{\rm
Im}~x_1} &=& 0\,, \label{4poles2}
\\
{\rm Re}(x_1 - x_2)\left(\frac{\beta + \gamma}{|x_1 - x_2|^2} +
\frac{\beta - \gamma}{|x_1 - x^*_2|^2}\right) &=&
0\,.\label{4poles3}
\end{eqnarray}
\noindent It follows from Eq.~(\ref{4poles3}) that ${\rm Re}~x_1 =
{\rm Re}~x_2\,.$ This solution describes the ``confluence'' of
poles. Then the remaining Eqs.~(\ref{4poles1}), (\ref{4poles2})
give
\begin{eqnarray}
{\rm Im}~x_{1,2} = \frac{1}{\alpha} \left(1 + 2~\frac{\beta -
\gamma}{\beta + \gamma}\right) \left(1 \pm \sqrt{\frac{\beta +
\gamma}{2\beta}}\right) = \left(\theta^2 - \theta + 1\right)
\left(1 \pm \sqrt{\frac{2\theta}{\theta^2 + 1}}\right)\,.
\nonumber
\end{eqnarray}
\noindent The two-pole and four-pole solutions are compared in
Fig.~3 in the case of $\theta = 8$ and $x_0 = 0.$

\begin{figure}
\scalebox{0.7}{\includegraphics{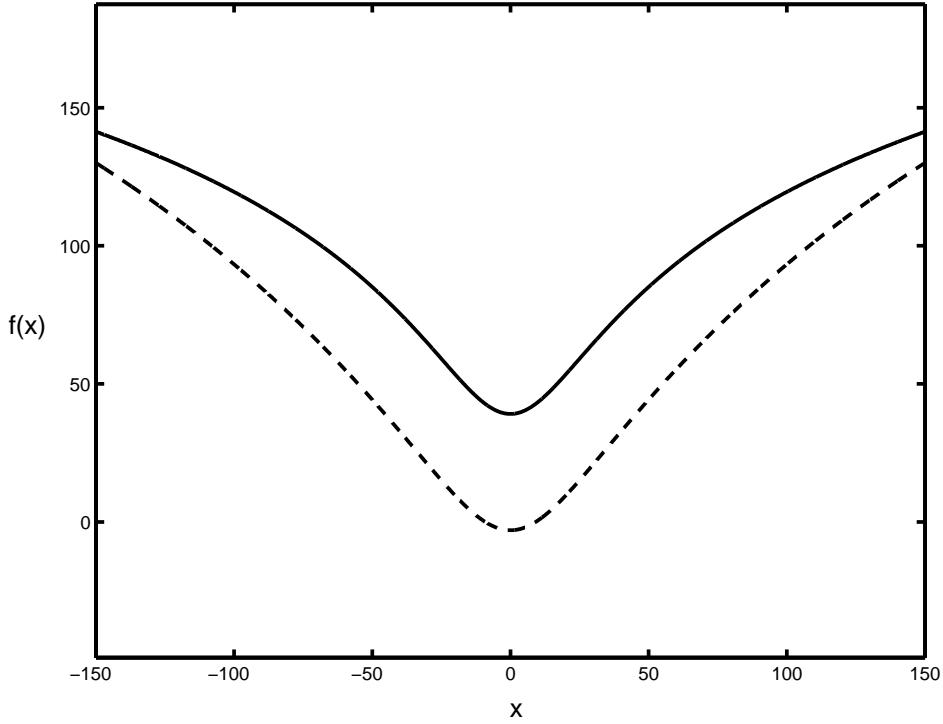}} \label{fig3}
\caption{Two-pole (full line) and four-pole (dashed line)
solutions of Eq.~(\ref{stationary1}) for $\theta = 8.$}
\end{figure}

The pole confluence is in fact a common property of the solutions
(\ref{poledec}). To see this, let us take the real part of the
equation with $k$ corresponding to the rightmost pole in the upper
half-plane. We have
\begin{eqnarray}
\sum\limits_{\genfrac{}{}{0pt}{}{l=1}{l\ne k}}^{2P} \left({\rm Re
}~x_k - {\rm Re}~x_l\right)\frac{\beta \pm \gamma}{|x_k - x_l|^2}
= 0\,. \nonumber
\end{eqnarray}
\noindent In view of the choice of $k,$ the left hand side is the
sum of non-negative terms. It can be zero only if ${\rm Re }~x_k -
{\rm Re}~x_l = 0$ for all $l.$

The question of which configuration is realized in the given
conditions requires carrying out the stability analysis of various
pole solutions, and can be solved, of course, only on the basis of
the general non-stationary Eq.~(\ref{nonstationary}). According to
the definition of $\mathfrak{f},$ such an analysis is to be
performed with respect to perturbations with wavelengths $\lambda
\sim L_0.$

\section{Discussion and conclusions}\label{conclude}

The large scale flame dynamics are independent of its local
cellular structure in zero order approximation with respect to the
flame front thickness. This is the main result of the work, proved
in Sec.~\ref{theorem}. The local flame corrugation only affects
the value of the normal velocity $U_{\rm f}$ changing it to
$\mathfrak{U}U_{\rm f},$ where $\mathfrak{U}>1$ describes increase
of the front length due to its wrinkling. In the scope of the thin
front model, $U_{\rm f}$ plays the role of an external parameter
specifying the characteristic velocity of the problem under
consideration. Thus, the overall effect of the local flame
structure on its large scale evolution amounts to a
renormalization of this parameter. For flames of practical
importance, the $\mathfrak{U}$-factor is about $1.3 - 1.5.$ In
fact, it is $\mathfrak{U}U_{\rm f},$ rather than $U_{\rm f},$
which is more convenient to measure experimentally, since the
measurement of $U_{\rm f}$ requires special facilities to suppress
development of the LD-instability, such as those used in
Ref.~\cite{clanet}.

The decoupling theorem allows one to avoid the difficult issues
arising in investigating flame dynamics at length scales of the
order $\lambda_{\rm c},$ and to go directly to scales
characterizing the problem in question. This is particularly
important in numerical simulations of the flame dynamics. The
computational grid should be chosen so as to well resolve the
flame cellular structure, which leaves a little space for
investigation of larger scales because of the limitations of
computational facilities.

The decoupling theorem also opens the way for analytical
investigation of the large scale flame dynamics. As an example,
the nonlinear development of the LD-instability in the presence of
the gravitational field was considered in Sec.~\ref{example},
where a weakly nonlinear non-stationary equation for the flame
front position was obtained [Eq.~(\ref{nonstationary})]. This
equation admits stationary solutions in the case of flame
propagation in the direction of the field, which means that the
gravitational field has a stabilizing overall effect in this case.
The resulting stationary flame configuration turns out to be
essentially non-periodic, and represents a symmetrical ``hump'' in
the direction of the flame propagation, with slowly decreasing
logarithmic ``tails.'' A complete investigation of the
non-stationary equation will be given elsewhere.

\end{document}